\documentclass[12pt,draftclsnofoot,onecolumn]{IEEEtran}
\usepackage{cite}
\usepackage{amssymb}
\usepackage[thmmarks]{ntheorem}
\usepackage[cmex10]{amsmath}
\allowdisplaybreaks[4]
\usepackage{algorithmic}
\usepackage{array}
\usepackage{url}
\usepackage{amsfonts}
\usepackage{CJK}
\usepackage{indentfirst}
\usepackage{amsmath}

\ifCLASSINFOpdf
  \usepackage[pdftex]{graphicx}
\else
  \usepackage[dvips]{graphicx}
\fi

\begin{document}
\title{\huge{Joint Transceiver and Power Splitter Design\\Over Two-Way
Relaying Channel\\with Lattice Codes and Energy Harvesting}}
\author{Zhigang Wen,~Shuai Wang, Chunxiao Fan and Weidong Xiang
\thanks{This work was previously published in IEEE Communications Letters, vol. 18, no. 11.
This version corrects some errors in the published letter.
More details can refer to \cite{12}.}
\thanks{Zhigang Wen, Shuai Wang, and Chunxiao Fan are with  the Beijing Key Laboratory of Work Safety Intelligent Monitoring, School of Electronic Engineering, Beijing University of Posts and Telecommunications, Beijing 100876, P.R.China (e-mail: zwen@bupt.edu.cn; comb0205@bupt.edu.cn; cxfan@bupt.edu.cn).}
\thanks{Weidong Xiang is with the Department of Electrical and Computer Engineering, University of Michigan-Dearborn,Dearborn, MI 48126, USA (e-mail:xwd@umich.edu).}

}

\maketitle
\begin{abstract}
This letter considers a compute-and-forward two-way relaying channel with simultaneous wireless information and power transfer.
Specifically, two single-antenna users exchange information via a multi-antenna relay station based on nested lattice codes.
Meanwhile, wireless energies flow from the relay to users for circuit consumption and uplink transmission.
Based on this model, an optimization problem is formulated to minimize the transmit power at relay, while guaranteeing the minimal transmission rate at each user.
To solve the problem, we propose an efficient iterative algorithm to jointly optimize the transmitter, receiver and power splitter, based on semi-definite relaxation and semi-definite programming.
Numerical results of relay transmission powers validate our analysis.
\end{abstract}


\IEEEpeerreviewmaketitle

\section{Introduction}

\IEEEPARstart{S}imultaneous wireless information and power transfer (SWIPT) refers to the transmission of powers to energy harvesters (EHs) and signals to information decoders (IDs) over the same electromagnetic waves \cite{1}.
Recently, SWIPT receives considerable attention in amplify-and-forward (AF) relay networks.
For example, when the relay nodes (RNs) are energy-constrained \cite{2,3,4,5}, SWIPT can enable RNs to harvest power for relaying signals.
In particular in \cite{2}, the RNs can apply power splitting (PS) or time switching for AF relaying.
Furthermore, multiple RNs can adopt SWIPT for tranmission and the outage probability of such a network is analyzed in \cite{3,4}.
On the other hand, when user terminals (UTs) are energy constrained \cite{6,7}, SWIPT can support UTs to collect power for transmitting signals.
This type of SWIPT relay networks is discussed in \cite{6} for AF one-way relaying channel and in \cite{7} for AF two-way relaying channel (TWRC), both of which focus on the optimization of beam-forming (BF) design.

However, in the context of TWRC, it is well known that AF relaying suffers from the noise amplification.
Thus a more promising way is to adopt compute-and-forward (CoF) scheme based on lattice codes (LC), since LC-CoF can reach the cut-set bound of TWRC within 1/2 bits [8].
In this sense, the study of SWIPT in TWRC with LC-CoF is necessary.
While the authors of [9] discuss such system with single-antenna relay,
this paper takes a step further to consider a multi-antenna TWRC with LC-CoF and PS-SWIPT.
In particular, UTs are equipped with energy harvesters, and RN acts as an information forwarder as well as power supplier.
For this system, we apply optimization techniques to guarantee the required data rates for UTs while minimizing the relay transmit power.

The contribution of the letter is two-fold.
Firstly, we propose a two-way transmission scheme with LC-CoF and PS-SWIPT in multi-antenna relay networks.
By analyzing the expression of achievable rate region, a power minimization problem with multiple variables is therefore formulated.
Secondly, an iterative joint algorithm is proposed to response to it.
The sub-problems in each iteration are further solved using semi-definite relaxation (SDR) and semi-definite programming (SDP).
Numerical results uncover good performances with low transmit power at nominal data rates.

\section{Preliminaries and System Model}

\subsection{Preliminaries on Lattice Codes}
Algebraically, an $n$-dimension lattice $\mathrm{\Lambda}$ is discrete subgroup in the Euclidean space $\mathbb{R}^n$ under vector addition. Thus, if $\mathbf{\mathbf{\lambda_1}},\mathbf{\mathbf{\lambda_2}}$ are in $\mathrm{\Lambda}$, their sum and difference are also in $\mathrm{\Lambda}$. The zero vector is always an element in a lattice.
Below, we provide fundamental concepts for lattice codes.

\emph{Definition 1}: A lattice quantizer $Q_{\mathrm{\Lambda}}$ maps a point $\mathbf{p}\in\mathbb{R}^n$ to a nearest point in $\mathrm{\Lambda}$:
\begin{equation}\label{2}
Q_{\mathrm{\Lambda}}(\mathbf{p})=\mathop{\rm{arg~min}}_{\mathbf{\lambda}\in\mathrm{\Lambda}}||\mathbf{p}-\mathbf{\lambda}||,
\end{equation}
and the operation $\rm{mod}$ is defined as: $\mathbf{p}~\mathrm{mod}~\mathrm{\Lambda}= \mathbf{p}-Q_{\mathrm{\Lambda}}(\mathbf{p})$.

\emph{Definition 2}:
The Voronoi Region of $\mathrm{\Lambda}_i$ is defined as all points that are closet to zero vector:
\begin{equation}\label{2}
V_i=\{\mathbf{p}:Q_{\mathrm{\Lambda}_i}(\mathbf{p})=0\}.
\end{equation}

\emph{Definition 3}:
The second moment of a lattice ¦« is defined as the second moment per dimension of a uniform distribution over the fundamental Voronoi region:
\begin{equation}\label{2}
\sigma^2(\Lambda_i)=\frac{1}{n\mathrm{Vol}(V_i)}\int_{V_i}||\mathbf{x}||^2d\mathbf{x},
\end{equation}
where $\mathrm{Vol}(V_i)$ is the volume of $V_i$.

\emph{Definition 4}:
A nested lattice can be defined as two $n$-dimension lattices $\mathrm{\Lambda}_i$ and $\mathrm{\Lambda}$ that form partition chain, i.e. $\mathrm{\Lambda}_i\subseteq\mathrm{\Lambda}$.

\emph{Definition 5}:
The lattice codebook $L_i$ use $\mathrm{\Lambda}$ as codewords and Voronoi Region $V_i$ as a shaping region:
\begin{equation}\label{2}
L_i=\{\mathrm{\Lambda}~\mathrm{mod}~\mathrm{\Lambda}_i\}=\{\mathrm{\Lambda}\cap V_i \}.
\end{equation}

\subsection{System Model and Transmission Scheme}
\begin{figure}[!t]
\centering
\includegraphics[width=120mm]{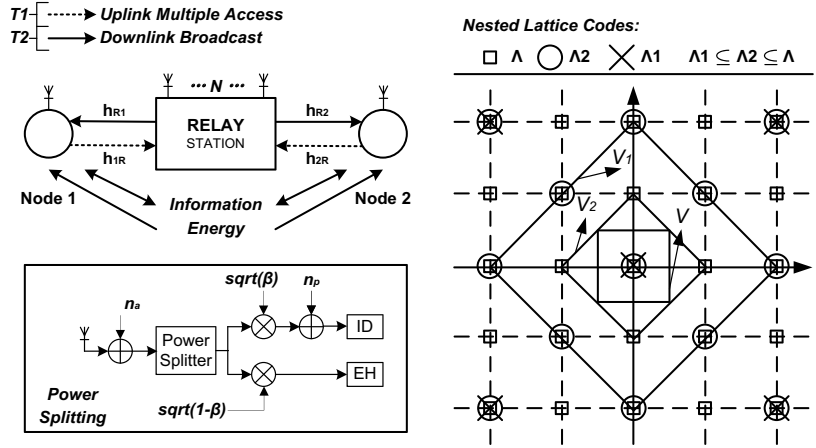}
\caption{System model of TWRC with LC-CoF and SWIPT}
\label{fig_sim}
\end{figure}

\setcounter{secnumdepth}{4}As shown in Fig.1, a TWRC system consists of two single-antenna nodes and a multi-antenna relay station, where $N$ antennas at the relay station are used to reduce the propagation loss by steering beams towards intended directions \cite{1}.
For this system, the uplink channel vector can be modeled as $\mathbf{h}_{i}\in \mathbb{C}^{N\times 1}$, which remains unchanged for the duration of $T$, while the downlink channel vector can be obtained with reciprocity.
All the noises are assumed to follow $\mathcal{CN}(0,\sigma^2)$.
The transmission contains two stages, and the details of each stage are given below.

In the first \emph{multiple access} (MA) stage,
given a $T$-dimension three chain doubly nested lattice $\mathrm{\Lambda}_1\subseteq\mathrm{\Lambda}_2\subseteq\mathrm{\Lambda}$,
we generate lattice codebook $L_i=\{\mathrm{\Lambda}\cap V_i \}$, and then map source codes $\mathbf{c}_i\in \{1,2...2^{2TR_i}\}$ into $\mathbf{w}_i\in L_i$, where $R_i$ is the code rate for $i$.
Adding dither random vectors $\mathbf{u}_i\in \mathbb{C}^{1\times T}$, source terminals transmit:
\begin{eqnarray}\label{x}
&\mathbf{x}_i={(\mathbf{g}\mathbf{h}_i)}^{-1}[(\mathbf{w}_i+\mathbf{u}_i)\mathrm{mod}~\Lambda_i],~~i=1,2.
\end{eqnarray}
Since the transmit symbol $\mathbf{x}_i\in \mathbb{C}^{1\times T}$ satisfies $\frac{1}{T}E(||\mathbf{x}_i||^2)= P_i$, the second moments of $\Lambda_i$ are $\sigma^2(\Lambda_i)=P_i|\mathbf{g}\mathbf{h}_i|^2$.
Based on the above scheme, the received signal $\mathbf{y}_r$ at relay on a single sub-carrier can be modeled as:
\begin{eqnarray}\label{1}
&\mathcal{MA}:~\mathbf{y}_r=\mathbf{g}(\mathbf{h}_1\mathbf{x}_1+\mathbf{h}_2\mathbf{x}_2+\mathbf{N}_r),
\end{eqnarray}
where $\mathbf{g}\in \mathbb{C}^{1\times N}$ with $||\mathbf{g}||^2=1$ is the receiver vector, and $\mathbf{N}_r\in \mathbb{C}^{N\times T}$ is the noise at relay.
Using the equation \eqref{x}, the received signal $\mathbf{y}_r$ becomes:
\begin{eqnarray}\label{1}
&\mathbf{y}_r=\mathop{\sum}_{i=1}^2(\mathbf{w}_i+\mathbf{u}_i)\mathrm{mod}~\Lambda_i+\mathbf{g}\mathbf{N}_r,
\end{eqnarray}
where noise power $\frac{1}{T}E[||\mathbf{g}\mathbf{N}_r||^2]=\sigma^2$.
Based on the lattice scheme in \cite{8}, the relay can then compute $(\alpha\mathbf{y}_r-\Sigma_{i=1}^2\mathbf{u}_i)\mathrm{mod}~\Lambda_1$ to recover $\mathbf{t}$,
where $\mathbf{t}$ is the target estimate and $\alpha$ is the MMSE coefficient:
\begin{align}\label{1}
&\mathbf{t}=[\mathbf{w}_1+\mathbf{w}_2-Q_{\Lambda_2}(\mathbf{w}_2+\mathbf{u}_2)]\mathrm{mod}~\Lambda_1 \nonumber \\
&\alpha=\frac{P_1|\mathbf{g}\mathbf{h}_1|^2+P_2|\mathbf{g}\mathbf{h}_2|^2}{P_1|\mathbf{g}\mathbf{h}_1|^2+P_2|\mathbf{g}\mathbf{h}_2|^2+\sigma^2}.
\end{align}

In the second \emph{broadcast} (BC) stage, the relay can use another lattice codebook $L_r$ to map the $\mathbf{t}$ to a symbol $\mathbf{x}_r(\mathbf{t})\in L_r$ where $\frac{1}{T}E(||\mathbf{x}_r||^2)=P_r$ and $P_r$ is the transmission power at $r$ to minimize. Therefore,
\begin{eqnarray}\label{1}
&\mathcal{BC}:~\mathbf{y}_i=\mathbf{h}^{\mathrm{T}}_{i}\mathbf{f}\mathbf{x}_r+\mathbf{n}_i,~~i=1,2,
\end{eqnarray}
where $\mathbf{f}\in \mathbb{C}^{N\times 1}$ with $||\mathbf{f}||^2=1$ is the beam-forming vector.
The received signal at the $i^{\mathrm{th}}$ node in BC is divided by the power splitting factor $\beta_i$ into two branches, of which one stream to EH and the other to ID.

At the EH side, the harvested energy is given by $\eta(1-\beta_i)P_r|\mathbf{h}^{\mathrm{T}}_{i}\mathbf{f}|^2$.
Since the two nodes are both powered \emph{only} by relay, the uplink transmit power of the $i^{\mathrm{th}}$ node is given by:
\begin{equation}
P_i=\eta(1-\beta_i)P_r|\mathbf{h}^{\mathrm{T}}_{i}\mathbf{f}|^2-2P_c,
\end{equation}
where $0<\eta<1$ is power conversion efficiency and $P_c$ is the circuit power in a symbol time.
At the ID side, the signal is given by
\begin{equation}\label{1}
\widetilde{\mathbf{y}}_i=\sqrt{\beta_i}\mathbf{h}^{\mathrm{T}}_{i}\mathbf{f}\mathbf{x}_r+\sqrt{\beta_i}\mathbf{n}_a+\mathbf{n}_p,
\end{equation}
where $\mathbf{n}_a,\mathbf{n}_p$ are noises before and after the power splitter, with power $\sigma^2_a,\sigma^2_p$ satisfying $\sigma^2_a+\sigma^2_p=\sigma^2$.

Denonting $R_{a,b}$ as the transmission rate from node $a$ to $b$, as $T\rightarrow \infty$,
the achievable rate region is formulated as below \cite{8}:
\begin{align}\label{1}
&R_{i,r}<\frac{1}{2}{\Big[\mathrm{log}\Big(\gamma_i+\frac{P_i|\mathbf{g}\mathbf{h}_i|^2}{\sigma^2}\Big)\Big]}^+
\nonumber \\
&R_{r,i}<\frac{1}{2}\mathrm{log}\Big(1+\frac{\beta_iP_r|\mathbf{h}^{\mathrm{T}}_{i}\mathbf{f}|^2}{\beta_i\sigma^2_a+\sigma^2_p}\Big)\approx \frac{1}{2}\mathrm{log}\Big(1+\frac{\beta_iP_r|\mathbf{h}^{\mathrm{T}}_{i}\mathbf{f}|^2}{\sigma^2}\Big),
\end{align}
where $\sigma_a\ll \sigma_p$ and $\gamma_i$ is the ratio of the second moments for $\Lambda_i$:
\begin{equation}
\gamma_i=\frac{\sigma^2(\Lambda_i)}{\sigma^2(\Lambda_1)+\sigma^2(\Lambda_2)}=\frac{P_i|\mathbf{g}\mathbf{h}_i|^2}
{P_1|\mathbf{g}\mathbf{h}_1|^2+P_2|\mathbf{g}\mathbf{h}_2|^2}.
\end{equation}

\section{Proposed Joint Transceiver and Power Splitter Design}
\subsection{Problem Formulation}

In quality of service (QoS) communication system, the transmission rate at each terminal should be guaranteed above a minimum value;
otherwise outage may happen \cite{10}.
Assume the required data rate sent from the $i^{\mathrm{th}}$ terminal is $\overline{R}_{i}$.
Since
$R_{1}=\mathrm{min}(R_{1,r},R_{r,2})$ and $R_{2}=\mathrm{min}(R_{2,r},R_{r,1})$,
an optimization problem can be formulated as:
\begin{align}\label{2}
&~\mathop{\mathrm{min}}_{P_r,\mathbf{f},\mathbf{g},\{P_i,\beta_i,\gamma_i\}}~~P_r \nonumber\\
&~\mathrm{s.t.:}~~\gamma_i+P_i\frac{|\mathbf{g}\mathbf{h}_i|^2}{\sigma^2}\geq \theta_{i,r},~\forall i \nonumber\\
&~~~~~~~~~1+\frac{\beta_iP_r|\mathbf{h}^{\mathrm{T}}_{i}\mathbf{f}|^2}{\sigma^2}\geq \theta_{r,i},~\forall i \nonumber
\\
&~~~~~~~~~
P_i=\eta(1-\beta_i)P_r|\mathbf{h}^{\mathrm{T}}_{i}\mathbf{f}|^2-2P_c,~~\forall i\nonumber
\\
&~~~~~~~~~\beta_i\in [0,1],~~\forall i\nonumber
\\
&~~~~~~~~~
||\mathbf{f}||^2=1,~||\mathbf{g}||^2=1 \nonumber
\\
&~~~~~~~~~
\gamma_i=\frac{P_i|\mathbf{g}\mathbf{h}_i|^2}
{P_1|\mathbf{g}\mathbf{h}_1|^2+P_2|\mathbf{g}\mathbf{h}_2|^2},~~\forall i,
\end{align}
where $\theta_{i,r}=2^{2\overline{R}_{i}}$ and $\theta_{r,i}=2^{2\overline{R}_{3-i}}$.

Due to the nonlinear equality of $\gamma_i$, the above optimization problem is difficult.
To this end, we relax the last constraint and set the value of $\gamma_i=0$ due to $\gamma_i<1\ll \theta_{i,r}$, which would yield an upper bound on the optimal $P^*_r$.
Then combining the first four constraints, we arrive at:
\begin{align}\label{2}
&0\leq\frac{\sigma^2(\theta_{r,i}-1)}{P_r|\mathbf{h}_i^{\mathrm{T}}\mathbf{f}|^2}\leq 1-\frac{\sigma^2\theta_{i,r}}{\eta P_r|\mathbf{h}_i^{\mathrm{T}}\mathbf{f}|^2|\mathbf{g}\mathbf{h}_i|^2}
-\frac{2P_c}{\eta P_r|\mathbf{h}_i^{\mathrm{T}}\mathbf{f}|^2}\leq 1
,~\forall i.
\end{align}
Thus the original optimization problem can be reformulated as:
\begin{align}\label{P2}
&~\mathop{\mathrm{min}}_{P_r,\mathbf{f},\mathbf{g}}~~~P_r \nonumber \\
&~\mathrm{s.t.:}~~P_r\mathbf{f}^{\dag}\mathbf{h}_i^*\mathbf{h}_i^{\mathrm{T}}\mathbf{f}\geq
\frac{\sigma^2\theta_{i,r}}{\eta\mathbf{g}\mathbf{h}_i\mathbf{h}^{\dag}_i\mathbf{g}^{\dag}}+\sigma^2(\theta_{r,i}-1)+\frac{2P_c}{\eta},~\forall i \nonumber \\
&~~~~~~~~~\mathbf{f}^{\dag}\mathbf{f}=1,~\mathbf{g}\mathbf{g}^{\dag}=1.
\end{align}
Now the problem is only related to the relay power $P_r$, transmitter $\mathbf{f}$ and receiver $\mathbf{g}$, which can be optimized iteratively as follows.

\subsection{Beam-forming Vector Design at Relay}
Consider the \emph{sub-problem 1} of problem \eqref{P2} to find $P_r$ and $\mathbf{f}$ given $\mathbf{g}$.
Let $g_i=|\mathbf{g}\mathbf{h}_{i}|^2$ be effective uplink channel gains.
Reformulate the problem by doing the following transformations:
\begin{align}
\mathbf{F}=P_r\mathbf{f}\mathbf{f}^{\dag}\in \mathbb{C}^{N\times N}.
\end{align}
Then since $P_r=\mathrm{Tr}\{P_r\mathbf{f}\mathbf{f}^{\dag}\}=\mathrm{Tr}\{\mathbf{F}\}$ and $
P_r\mathbf{f}^{\dag}\mathbf{A}_i\mathbf{f}=\mathrm{Tr}\{P_r\mathbf{f}^{\dag}\mathbf{A}_i\mathbf{f}\}=\mathrm{Tr}\{\mathbf{A}_i\mathbf{F}\}$,
we arrive at:
\begin{align}\label{2}
&~\mathop{\mathrm{max}}_{\mathbf{F}\succeq 0}~~\mathrm{Tr}(\mathbf{F})\nonumber \\
&~\mathrm{s.t.:}~~\mathrm{Tr}\{\mathbf{A}_i\mathbf{F}\}\geq a_i,~\forall i=1,2
\nonumber\\
&~~~~~~~~
\mathrm{Rank}\{\mathbf{F}\}=1,
\end{align}
where the parameters
\begin{equation}
\left\{
\begin{aligned}
&
\mathbf{A}_i=\mathbf{h}_i^*\mathbf{h}_i^{\mathrm{T}}\\
&a_i=\frac{\sigma^2\theta_{i,r}}{\eta g_i}+\sigma^2(\theta_{r,i}-1)+\frac{2P_c}{\eta}\\
\end{aligned}
.
\right.\nonumber
\end{equation}

Problem (18) is still non-convex due the rank constraint $\mathrm{Rank}(\mathbf{F})=1$.
By using SDR to relax the non-convex constraint, the resultant SDP problem can be solved by CVX, a Matlab software package for solving convex problems \cite{11}.
Furthermore, since problem (18) has only two constraints on $\mathbf{F}$, there will always exist an optimal solution $\mathbf{F}^*$ to the SDR problem of (18) satisfying $\mathbf{F}^*\leq \sqrt{2}$, namely $\mathbf{F}^*$ is rank-one.
Therefore, the SDR will not change the problem.
With this rank-one solution $\mathbf{F}^*$, the singular value decomposition (SVD) of $\mathbf{F}^*$ is
\begin{align}
&\mathbf{F}^*=\mathbf{Z}\mathbf{\Gamma}\mathbf{Z}^{\dag},~\mathbf{\Gamma}=\mathrm{diag}(P_r,0,...,0),
\mathbf{Z}=[\mathbf{z}_1, ..., \mathbf{z}_N], \nonumber
\end{align}
and the projection is $\mathbf{f}^*=\mathbf{z}_1$.

\subsection{Combining Vector Design at Relay}

\setcounter{secnumdepth}{4}Next consider the \emph{sub-problem 2} of \eqref{P2} to find $P_r$ and $\mathbf{g}$ given $\mathbf{f}$.
Let $h_i=|\mathbf{h}_i^{\mathbf{T}}\mathbf{f}|^2$, and we arrive at an equivalent optimization problem after eliminating $P_r$
\begin{align}\label{2}
&\mathop{\mathrm{min}}_{\mathbf{g}\in\mathbb{C}^{1\times N}}~
\mathop{\mathrm{max}}_{i=1,2}\Big(\frac{\rho_i}{\mathbf{g}\mathbf{h}_i\mathbf{h}^{\dag}_i\mathbf{g}^{\dag}}+\mu_i\Big)
\nonumber\\
&~~~\mathrm{s.t.:}~~\mathbf{g}\mathbf{g}^{\dag}=1,
\end{align}
where the coefficients are given by:
\begin{equation}
\begin{aligned}
&\rho_i=\frac{\sigma^2\theta_{i,r}}{\eta h_i},~\mu_i=\frac{\sigma^2(\theta_{r,i}-1)+2P_c\eta^{-1}}{h_i}.
\end{aligned}
\end{equation}

The problem above contains quadratic functions.
To this end, change the variables as $\mathbf{G}=\mathbf{g}^{\dag}\mathbf{g}\in \mathbb{C}^{N\times N}$, and the problem can be transformed into
\begin{align}
&\mathop{\mathrm{min}}_{\mathbf{G}\succeq 0}~
\mathop{\mathrm{max}}_{i=1,2}\Big(\frac{\rho_i}{\mathrm{Tr}(\mathbf{h}_i\mathbf{h}^{\dag}_i\mathbf{G})}+\mu_i\Big)
\nonumber\\
&~~~\mathrm{s.t.:}~~\mathrm{Tr}(\mathbf{G})=1,\mathrm{Rank}\{\mathbf{G}\}=1.
\end{align}
Following similar reason under problem (18), the rank constraint $\mathrm{Rank}(\mathbf{G})=1$ can be dropped without changing the problem.
Then the relaxed problem is convex and can be solved by CVX.
With the obtained $\mathbf{G}^*$, the SVD of $\mathbf{G}^*$ is
\begin{align}
&\mathbf{G}^*=\mathbf{U}\mathbf{\Lambda}\mathbf{U}^{\dag}, \mathbf{\Lambda}=\mathrm{diag}(1,0,...,0),
\mathbf{U}=[\mathbf{u}_1, ..., \mathbf{u}_N],
\end{align}
and the projection is $\mathbf{g}^*=\mathbf{u}^{\dag}_1$.

\subsection{Summary of Algorithm}

With the optimal $P^*_r,\mathbf{f}^*,\mathbf{g}^*$ to problem (16) being obtained, the PS ratio can be given by
\begin{align}
&\beta^*_i=\frac{1}{2}\Big(1+
\frac{\eta\sigma^2(\theta_{r,i}-1)-2P_c}{\eta P^*_r|\mathbf{h}_i^{\mathrm{T}}\mathbf{f}^*|^2}-\frac{\sigma^2\theta_{i,r}}
{\eta P^*_r|\mathbf{h}_i^{\mathrm{T}}\mathbf{f}^*|^2|\mathbf{g}^*\mathbf{h}_i|^2} \Big).
\end{align}
The total procedure is given in \emph{Algorithm 1}, which is fast-convergent requiring few iterations.

\emph{\underline{Algorithm 1:}} \emph{Joint transceiver-PS in TWRC with LC-CoF and SWIPT.}

1: \textbf{Initialize} $\gamma_i=0,\mathbf{g}=\sqrt{\frac{1}{N}}[1,...,1], \forall i$.

2: \textbf{Repeat}.

3: Calculate the optimal $P_r,\mathbf{f}$ given $\mathbf{g}$.

4: Calculate the optimal $P_r,\mathbf{g}$ given $\mathbf{f}$.

5: Until \textbf{convergence}.

\section{Numerical Results}

In this section we provide numerical results and comparisons.
We consider the number of antennas $N=4$ and power conversion efficiency $\eta=1$.
Then 100 random channels with Rayleigh fading coefficients $\sim \mathcal{CN}(0,1)$ are generated in total for Monte Carlo simulation\footnote{
The impact of pathloss is not considered here.
However, the detailed propagation modeling in SWIPT could be found in \cite{3,4}, and it does not affect our comparisions.}.
The compared schemes are:
\emph{1.} joint transceiver and PS design;
\emph{2.} joint BF and PS design with equal gain combining (EGC) receiver;
\emph{3.} joint receiver and PS design with equal gain BF;
\emph{4.} PS design with no transceiver design.

Fig. 2 provides the transmission power versus SNR.
As SNR increases, the required transmission power decreases.
The proposed joint transceiver-PS design achieves the best performance,
while PS design without transceiver design is the worst.
The second best scheme is joint BF-PS design with $9\mathrm{dB}$ loss compared to the proposed method.
Notice that joint BF and PS design outperforms joint receiver and PS design under the same circumstance.

\begin{figure}[!t]
\centering
\includegraphics[width=80mm]{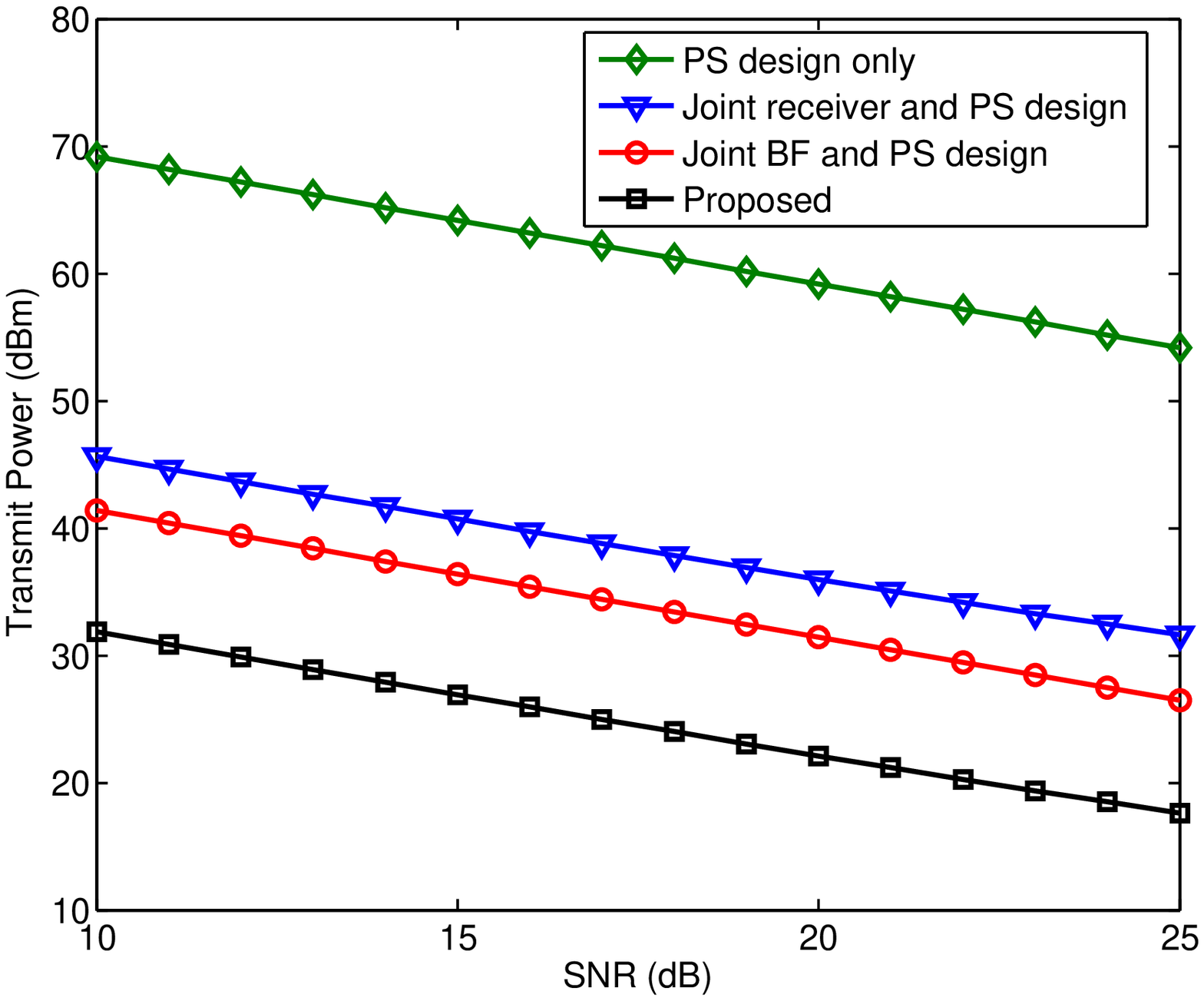}
\caption{Transmission power versus SNR.
$P_c=10\mathrm{dBm},\overline{R}_{1}=\overline{R}_{2}=2\mathrm{bps/Hz}$.}
\label{fig_sim}
\end{figure}

\begin{figure}[!t]
\centering
\includegraphics[width=80mm]{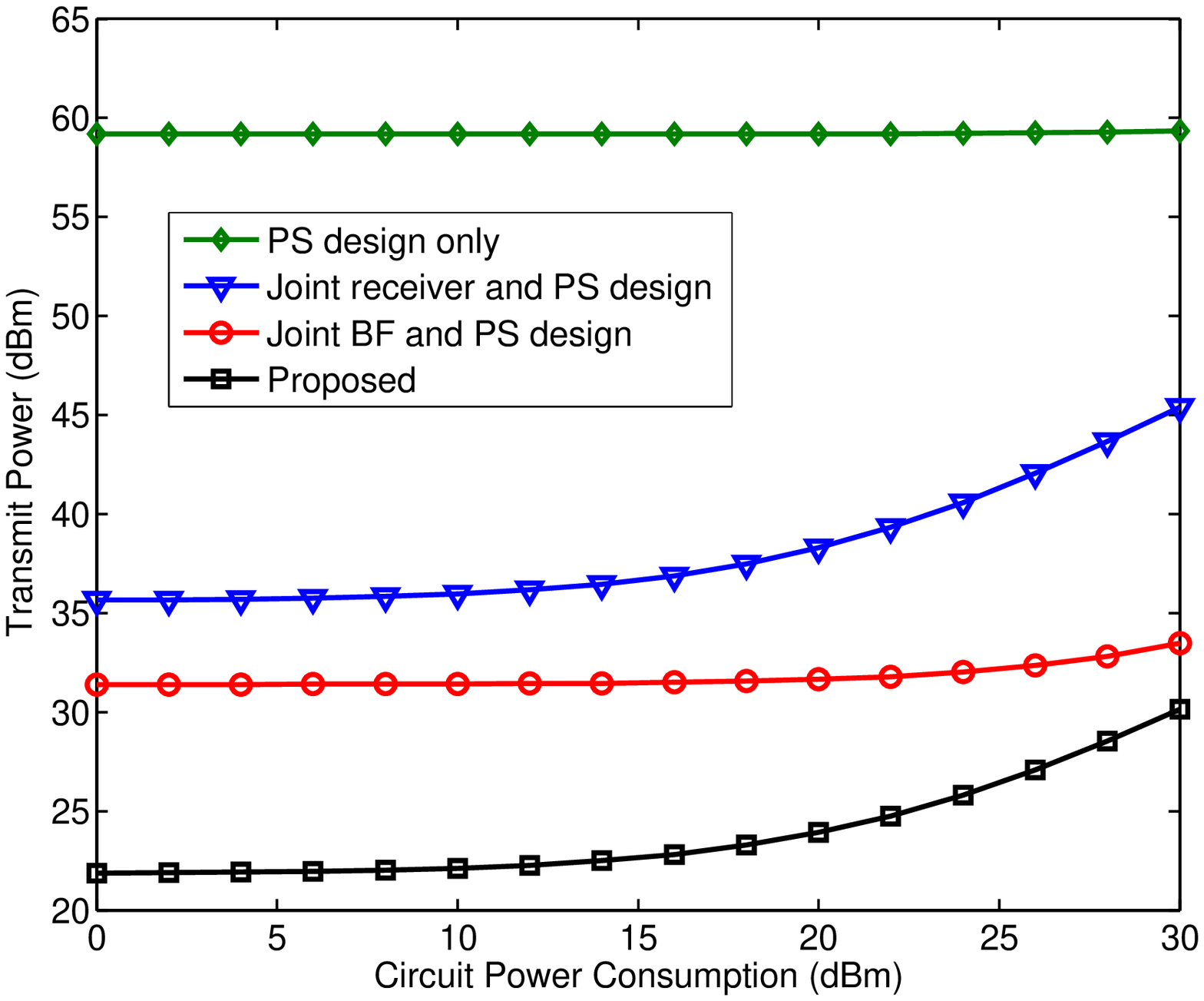}
\caption{Transmission power versus $P_c$.
$SNR=20\mathrm{dB},\overline{R}_{1}=\overline{R}_{2}=2\mathrm{bps/Hz}$.}
\label{fig_sim}
\end{figure}

Fig. 3 provides the transmission power versus circuit power consumption.
The power at relay increases when $P_c$ grows.
The proposed method still achieves the best performance,
but the gap between joint BF and PS design and the proposed method decreases when $P_c$ is larger.
Moreover, we still observe that joint BF and PS design outperforms joint receiver and PS design.
This indicates that in terms of the impacts on performance, beam-forming design$>$receiver design$>$PS design.

\section{Conclusion}

In this letter, we consider an SWIPT assisted TWRC system with CoF based on lattice codes.
Applying the method of SDR and SDP, we propose a joint design of transceiver and power splitter,
which can lead to low power cost while maintaining transmission rates at each terminal.
Numerical results of transmission power validate the proposed method, and the impacts of beam-former, receiver and power splitter are compared.
Future work may consider multi-antenna users and imperfect channel state information.

\ifCLASSOPTIONcaptionsoff
  \newpage
\fi

\end{document}